\documentclass[twocolumn,10pt]{ms}

\usepackage{epsfig} 

\title{SPH modeling of natural convection around a heated horizontal cylinder: A
       comparison with experiments}

\author{F. Arag\'on, J. E. V. Guzm\'an
    \affiliation{
	Instituto de Ingenier\'{\i}a UNAM\\ 
	Circuito Escolar S/N\\
	Ciudad Universitaria, Coyoac\'an\\
	04510 Ciudad de M\'exico, Mexico\\
        Email: micme2003@yahoo.com.mx\\
	Email: jguzmanv@iingen.unam.mx
    }	
}
\author{C. E. Alvarado-Rodr\'{\i}guez
    \affiliation{
	Direcci\'on de C\'atedras CONACYT\\
	Av. Insurgentes Sur 1582,\\ 
	Cr\'edito Constructor, Benito Ju\'arez\\
	03940 Ciudad de M\'exico, Mexico\\
        Email: carlos.alvarado@conacyt.mx
    }
}
\author{L. Di G. Sigalotti\thanks{Corresponding author.}
    \affiliation{
	Departamento de Ciencias B\'asicas\\
	Universidad Aut\'onoma Metropolitana\\
	(UAM-A), Av. San Pablo 180\\
	02200 Ciudad de M\'exico, Mexico\\
        E.mail: leonardo.sigalotti@gmail.com
    }
}
\author{I. Carvajal-Mariscal\\
    \affiliation{
	Instituto Polit\'ecnico Nacional (IPN)\\
	ESIME UPALM, Av. IPN s/n\\
	07738 Ciudad de M\'exico, Mexico\\
        E-mail: icarvajal@ipn.mx
    }
}
\author{J. Klapp\\
    \affiliation{
	Instituto Nacional de Investigaciones\\
	Nucleares, (ININ)\\ 
	Carretera M\'exico-Toluca km. 36.5\\
	La Marquesa, 52750 Ocoyoacac\\
	Estado de M\'exico, Mexico\\
        E-mail: jaime.klapp@inin.gob.mx
    }
}
\author{A. R. Uribe-Ram\'{\i}rez\\
    \affiliation{
        Departamento de Ingenier\'{\i}a Qu\'{\i}mica\\
	DCNyE, Universidad de Guanajuato\\ 
	Noria Alta S/N, 36000 Guanajuato\\ 
	Guanajuato, Mexico\\
        E-mail: agustin@ugto.mx
    }
}

\begin{document}

\maketitle    

\begin{abstract}
{\it An experimental and numerical smoothed particle hydrodynamics (SPH) analysis was
performed for the convective flow arising from a horizontal, thin cylindrical
heat source enclosed in a glycerin-filled, slender enclosure at low Rayleigh
numbers ($1.18\leq {\rm Ra}\leq 242$). Both the experiments and the SPH calculations
were performed for positive ($0.1\leq\Delta T\leq 10$ K) and negative
($-10\leq\Delta T\leq -0.1$ K) temperature differences between the source and the
surrounding fluid. In all cases a pair of steady, counter-rotating vortices is formed,
accompanied by a plume of vertically ascending flow just above the source for $\Delta T>0$
and a vertically descending flow just below the source for $\Delta T<0$. The maximum flow
velocities always occur within the ascending/descending plumes. The SPH predictions are
found to match the experimental observations acceptably well with root-mean-square
errors in the velocity profiles of the order of $\sim 10^{-5}$ m s$^{-1}$. The fact that
the SPH method is able to reveal the detailed features of the flow phenomenon demonstrates
the correctness of the approach.}
\end{abstract}


\begin{nomenclature}
\entry{${\rm Ra}$}{Rayleigh number.}
\entry{${\rm Pr}$}{Prandtl number.}
\entry{$H$}{Height of test cell, m.}
\entry{$L$}{Length of test cell, m.}
\entry{$W$}{Width of test cell, m.}
\entry{$D$}{Diameter of cylindical heat source, m.}
\entry{$m$}{Mass, kg.}
\entry{$\rho$}{Mass density, kg m$^{-3}$.}
\entry{${\bf v}$}{Fluid velocity vector, m s$^{-1}$.}
\entry{$T$}{Fluid temperature, K.}
\entry{$p$}{Pressure, Pa.}
\entry{${\bf F}$}{Body force, m s$^{-2}$.}
\entry{${\bf g}$}{Gravitational acceleration, m s$^{-2}$.}
\entry{$c$}{Sound speed, m s$^{-1}$.}
\entry{${\bf x}$}{Position vector, m.}
\entry{$t$}{Time, s.}
\entry{$x$}{Horizontal coordinate, m.}
\entry{$y$}{Vertical coordinate, m.}
\entry{$\nu$}{Kinematic viscosity, m$^{2}$ s$^{-1}$.}
\entry{$\eta$}{Shear viscosity, kg m$^{-1}$ s$^{-1}$.}
\entry{$\kappa$}{Thermal conductivity, W m$^{-1}$ K$^{-1}$.}
\entry{$c_{p}$}{Specific heat at constant pressure, J kg$^{-1}$ K$^{-1}$.}
\entry{$\gamma$}{Adiabatic exponent.}
\entry{$\beta$}{Thermal expansion coefficient, K$^{-1}$.} 
\entry{$\alpha$}{Thermal diffusivity, m$^{2}$ s$^{-1}$.}
\entry{$\Delta T$}{Temperature difference, K.}
\entry{$h$}{Smoothing length, mm.}
\entry{$\Delta t$}{Time step, s.}
\entry{$N$}{Total number of particles.}
\entry{$n$}{Number of neighboring particles.}
\entry{$a,b$}{Particle indices.}
\entry{$\rho _{\rm r}$}{Reference density, kg m$^{-3}$.}
\entry{$T_{\rm r}$}{Reference temperature, K.}
\entry{$T_{s}$}{Cylindrical source temperature, K.}
\entry{$T_{w}$}{Cell wall temperature, $^{\circ}$C.}
\entry{${\bf v}_{s}$}{Fluid velocity at the surface of the heat source, m s$^{-1}$.}
\entry{${\bf v}_{w}$}{Fluid velocity at the cell walls, m s$^{-1}$.}
\entry{$0$}{Reference index.}
\end{nomenclature}

\section{Introduction}

Heat transfer by natural convection in an enclosed space has received increasing
attention in the past years because of its importance in many engineering applications.
In particular, natural convective heat transfer around a single horizontal cylinder and
from horizontal tube arrays in cavities has direct applications in the design of heat
exchangers, nuclear reactors, solar heaters, radiators, thermal storage systems as well
as in the cooling of electronic components and the transport of oil through pipelines
below the water surface or at the sea floor. These applications have motivated for more
than 50 years a considerable body of research on their flow and heat transfer
characteristics. In particular, recent experimental and numerical analyses of 
heat transfer and flow characteristics of the air-water cross flow in cambered ducts have 
been addressed to raise the efficiency of heat exchangers \cite{Zhang2020}. 

Square and rectangular heated cavities have been widely explored both experimentally and
theoretically due to their simpler geometry. The main concerns of these
studies were first directed to laminar and turbulent natural convection around heated,
horizontal cylinders. At small Rayleigh numbers (Ra $<10^{4}$), the heat transfer from a
horizontal cylinder behaves like a line heat source \cite{Deschamps1994,Lauriat1994,
Linan1998,Cesini1999}. For larger Rayleigh numbers (i.e., $10^{4}\leq {\rm Ra}\leq 10^{8}$),
the flow forms a laminar boundary layer around the cylinder \cite{Kuehn1980,Farouk1981,
Fujii1982,Karim1986,Saitoh1993,Atmane2003,Angeli2008}, while at even higher Rayleigh numbers,
the motion of the fluid due to buoyancy is turbulent \cite{Kitamura1999,Misumi2003,
Reymond2008}. However, most of the research in this area has focused on laminar convective
flows, which are the simplest and the most common cases in natural convection. Earlier
numerical analyses by Kuehn and Goldstein \cite{Kuehn1980}, Farouk and G\"u\c{c}eri
\cite{Farouk1981} and Saitoh {\it et al.} \cite{Saitoh1993} have all shown that heat transfer
is maximum at the bottom of the cylinder and decreases towards the top of the cylinder. This
was experimentally assessed by Cesini {\it et al.} \cite{Cesini1999}, who also observed that
for Ra $>10^{5}$, oscillations of the heat flux at the surface of the cylinder appeared.
Moreover, Karim {\it et al.} \cite{Karim1986} performed experiments to study the effects of
horizontal confinement on heat transfer around a cylinder for $10^{3}\leq {\rm Ra}\leq 10^{5}$,
finding that the heat flux around the cylinder increases with decreasing distance between the
cylinder and the enclosure wall. For similar experiments, Cesini {\it et al.} \cite{Cesini1999}
found that the heat flux reaches a maximum for an optimal wall-to-wall distance of 2.9 times
the diameter of the cylinder. In contrast, when the flow medium around the cylinder
is vertically confined a complex combination of effects can be produced. For example, experiments
conducted by Koizumi and Hosokawa \cite{Koizumi1996} for vertical confinement of natural
convection around a hot horizontal cylinder by both conductive and adiabatic flat ceilings has
revealed chaotic and oscillatory movements of the air above the cylinder. Further experiments in
water by replacing the flat ceiling with a free-surface boundary were performed by Atmane
{\it et al.} \cite{Atmane2003}. They found that the primary effect of the vertical confinement was
an increase of the heat flux on the upper part of the cylinder for given separation distances from
the free water surface, which is related to the large-scale oscillation of the thermal plume.

A number of numerical analyses have been performed to study natural convection from horizontal
cylinders enclosed in square and rectangular cavities, using spectral element
\cite{Ghaddar1992,Ghaddar1994}, finite-difference \cite{De2006}, finite-element
\cite{Lauriat1994,Cesini1999}, and finite-volume \cite{Angeli2008,Kim2008,Adnani2016} methods. 
Articles dealing with Smoothed Particle Hydrodynamics (SPH) simulations of natural convection 
are not abundant in the literature. Extensions of SPH to fluid problems with heat transfer were 
first implemented by Cleary \cite{Cleary1998} and Cleary and Monaghan \cite{Cleary1999} under 
the Boussinesq approximation. Natural convection from a heated cylinder in a square enclosure 
was subsequently modeled by Moballa {\it et al.} \cite{Moballa2013} using an incompressible SPH 
approach. Their results were in good agreement with those from previous analyses based on 
conventional methods. SPH simulations with a non-Boussinesq formulation were also presented by 
Szewc {\it et al.} \cite{Szwec2011} to study natural convection under large density variations 
due to high temperature gradients. More recently, SPH simulations of natural convection in a
horizontal concentric annulus and in square cavities were reported by Yang and Kong 
\cite{Yang2019} and Garoosi and Shakibaeinia \cite{Garoosi2020a}, respectively. Further
SPH calculations of natural convection heat transfer in a differentially heated cavity were
reported by Garoosi and Shakibaeinia \cite{Garoosi2020b}. On the other hand, both lid-driven 
cavity and buoyancy-driven cavity SPH simulations were carried out by Hopp-Hirschler 
{\it et al.} \cite{Hopp2018a}, while forced convection heat transfer around a horizontal 
cylinder was also investigated using a weakly compressible SPH (WCSPH) scheme by Nasiri 
{\it et al.} \cite{Nasiri2019}. SPH studies of buoyancy-driven flow were recently presented 
by Hopp-Hirschler {\it et al.} \cite{Hopp2018b}, where the formation of viscous fingers is 
modeled in miscible and immiscible systems.

The thermal interaction between a cylindrical source and a rectangular enclosure was first
investigated numerically by Ghaddar \cite{Ghaddar1992}, who predicted flow patterns and
heat transfer rates for air over a wide range of Rayleigh numbers. In a more recent work,
De and Dalal \cite{De2006} studied numerically the effects of varying the cavity aspect ratio
on the flow pattern with the aid of finite-difference techniques. The present study will
further extend the capabilities of the SPH method to simulate natural convection heat transfer
problems. In particular, a weakly compressible SPH (WCSPH) formulation \cite{Becker2007}
based on the Boussinesq approximation is employed to reproduce experimental results of the
natural convection flow around a horizontal cylinder of small diameter in a slender cavity of
square cross-section $L$ and width $W=0.05L$. The SPH method is a fully Lagrangian, mesh-free 
scheme for fluid-flow simulations, where fluid elements are represented by a set of discrete 
particles that characterize the flow attributes. It presents several advantages over traditional 
Eulerian techniques because it can exactly simulate pure advection and deal with complex 
geometries and boundaries. It must be remarked that most of the previous work with this 
kind of geometry has focused, to the best of our knowledge, on investigating the process with 
low viscosity fluids such as air and water \cite{Angeli2011,Fiscaletti2013}. However, certain 
applications are driven by thermo-hydraulic phenomena that involve fluids with significant 
viscosity differences. For example, in the oil industry, the production of heavy oil relies on 
fluid-based thermal treatments to enhance the mobility of the highly viscous fluids 
\cite{Riyi2012,Zhao2017}. Clearly, the results and conclusions obtained in previous studies 
cannot be directly extrapolated in such cases. In particular, the present case may be
representative of a single section of a heat exchanger used to reduce the viscosity of heavy
crude oil in storage tanks.  

The governing equations and the SPH methodology used are described in Section 2.
The experimental rig and procedure are given in Section 3, while Section 4 describes the
numerical model and the implementation of the boundary conditions. The validation of the
numerical scheme along with the experimental and numerical results are presented in Section
5. The predictions obtained from the SPH simulations are found to be in good agreement
with the experimental results. Finally, Section 6 contains the main conclusions.

\section{Numerical methodology}

\subsection{Governing equations}

The equations describing the fluid motion and heat transfer around a horizontal
cylindrical heat source are the continuity, the Navier-Stokes, and the energy balance
equations. In Lagrangian form these equations are:
\begin{eqnarray}
\frac{d\rho}{dt}&=&-\rho\nabla\cdot {\bf v},\\
\frac{d{\bf v}}{dt}&=&-\frac{1}{\rho}\nabla p+\nu\nabla ^{2}{\bf v}+{\bf F},\\
\frac{dT}{dt}&=&\frac{1}{\rho c_{p}}\nabla\cdot\left(\kappa\nabla T\right),
\end{eqnarray}
where $d/dt$ is the material time derivative, $\rho$ is the density, ${\bf v}$ is the fluid
velocity vector, $p$ is the pressure, $\nu =\eta /\rho$ is the kinematic viscosity coefficient
and $\eta$ is the dynamic viscosity of the fluid, $T$ is the temperature, $c_{p}$ is the specific
heat at constant pressure, and $\kappa$ is the thermal conductivity. Since at low 
${\rm Ra}$-values heat generation due to viscous dissipation is negligible \cite{Garoosi2020a},
only the thermal conduction term is retained in the energy balance equation (3). This is
true because the total entropy generation due to thermal dissipation is by far much more intense 
than that due to viscous dissipation at ${\rm Ra}<10^{3}$ \cite{Garoosi2020a}.
In the Boussinesq approximation, the body force ${\bf F}$ on the right-hand side of Eq. (2) is 
given by
\cite{Gray1976}
\begin{equation}
{\bf F}=-\beta\left(T-T_{\rm r}\right){\bf g},
\end{equation}
where ${\bf g}$ is the gravitational acceleration, $\beta$ is the coefficient of thermal
expansion, and $T_{\rm r}$ is a reference temperature. In most applications this is the fluid
temperature at room conditions. For natural convection flow in the Boussinesq approximation, the
two relevant dimensionless numbers are: the Prandtl number
\begin{equation}
{\rm Pr}=\frac{\nu}{\alpha}=\frac{c_{p}\eta}{\kappa},
\end{equation}
where $\alpha =\kappa /(\rho c_{p})$ is the thermal diffusivity, and the Rayleigh number
\begin{equation}
{\rm Ra}=\frac{g\beta D^{3}\Delta T}{\alpha\nu},
\end{equation}
where $D$ is a characteristic length scale and $\Delta T$ is the leading temperature
difference between the heat source ($T_{s}=T_{\rm bath}$) and the surrounding fluid. In
particular, the value of Ra quantifies the ratio of buoyancy over viscosity in the fluid
for a given Prandtl number.

Equations (1)-(3) are closed by a pressure-density relation of the form \cite{Batchelor1974}
\begin{equation}
p=B\left[\left(\frac{\rho}{\rho _{\rm r}}\right)^{\gamma}-1\right],
\end{equation}
where $B=c_{0}^{2}\rho _{\rm r}/\gamma$, $\rho _{\rm r}$ is a reference density, $\gamma =7$
for liquids and $\gamma =1.4$ for gases, and $c_{0}$ is a numerical speed of sound which is
chosen to be 10 times higher than the maximum fluid velocity in order to keep the density
fluctuations below 1\% and satisfy the incompressibility condition \cite{Monaghan1992}.
The reference density $\rho _{\rm r}$ in Eq. (7) is allowed to change with the temperature
according to the following expression
\begin{equation}
\rho _{\rm r}=\frac{\rho}{\left[1+\beta\left(T_{\rm f}-T_{0}\right)\right]},
\end{equation}
where $T_{\rm f}$ and $T_{0}$ are, respectively, the final and initial temperatures
in a numerical timestep $\Delta t$.

\subsection{SPH formulation}

Equations (1)-(3) are solved numerically using a variant of the open-source code DualSPHysics
\cite{Gomez2012}, which relies on SPH theory \cite{Monaghan1992,Liu2010}. The temporal
variation of the fluid density at the position ${\bf x}_{a}$ of particle $a$ is calculated
according to the SPH representation \cite{Monaghan1992}
\begin{equation}
\frac{d\rho _{a}}{dt}=\sum _{b=1}^{n}m_{b}{\bf v}_{ab}\cdot\nabla _{a}W_{ab},
\end{equation}
where ${\bf v}_{ab}={\bf v}_{a}-{\bf v}_{b}$, $W_{ab}=W(|{\bf x}_{ab}|,h)$ is the kernel
interpolation, ${\bf x}_{ab}={\bf x}_{a}-{\bf x}_{b}$, $h$ is the width of the kernel
(or smoothing length), and the sum is over $n$ neighbours around particle $a$. This form is
Galilean invariant and is particularly convenient in the presence of free surfaces or
density discontinuities. For the momentum equation (2), we use the following representation
\begin{eqnarray}
\frac{d{\bf v}_{a}}{dt}&=&-\sum _{b=1}^{n}m_{b}\left(\frac{p_{a}+p_{b}}
{\rho _{a}\rho _{b}}\right)\nabla _{a}W_{ab}\nonumber\\
&+&2\nu\sum _{b=1}^{n}m_{b}\frac{{\bf v}_{ab}}{{\hat\rho}_{ab}}
\frac{{\bf x}_{ab}\cdot\nabla _{a}W_{ab}}{|{\bf x}_{ab}|^{2}+\epsilon ^{2}}+{\bf g},
\end{eqnarray}
where ${\hat\rho}_{ab}=(\rho _{a}+\rho _{b})/2$ and $\epsilon ^{2}=0.01h^{2}$ is a
small correction factor used to avoid singularities when $|{\bf x}_{ab}|\ll 1$.
Following Cleary \cite{Cleary1998}, the SPH form of the energy equation is given by
\begin{equation}
\frac{dT_{a}}{dt}=\frac{2\kappa}{\rho _{a}c_{p}}\sum _{b=1}^{n}\frac{m_{b}}{\rho _{b}}
\frac{T_{ab}{\bf x}_{ab}\cdot\nabla _{a}W_{ab}}{|{\bf x}_{ab}|^{2}+\epsilon ^{2}},
\end{equation}
where $T_{ab}=T_{a}-T_{b}$. This form ensures that the heat flux is automatically continuous
across material interfaces. In its original form derived by Cleary \cite{Cleary1998}
with $\kappa\to 2\kappa _{a}\kappa _{b}/(\kappa _{a}+\kappa _{b})$, it allows multiple
materials with substantially different conductivities and specific heats to be accurately
modelled. In a Lagrangian frame of reference, the SPH particles are moved according to
\begin{equation}
\frac{d{\bf x}_{a}}{dt}={\bf v}_{a},
\end{equation}
which must be solved simultaneously with Eqs. (9)-(11).

Since direct evaluation of second-order derivatives of the kernel function is not
required, a low-order Wendland C$^{2}$ function is used as the interpolation kernel
\cite{Dehnen2012}
\begin{equation}
W(q,h)=\frac{7}{\pi h^{2}}\left(1-q\right)^{4}\left(1+4q\right),
\end{equation}
for $0\leq q<1$ and zero otherwise, where $q=|{\bf x}-{\bf x}^{\prime}|/h$. Wendland
functions have positive Fourier transforms and so they can support arbitrarily large
numbers of neighbours without suffering from a close pairing of particles
\cite{Dehnen2012,Zhu2015,Sigalotti2019}. In addition, Wendland functions are very reluctant
to allow for particle motion on a sub-resolution scale, and in contrast to most commonly
used kernels they maintain a very regular particle distribution, even in highly dynamical
tests as occurs in turbulent regimes \cite{Rosswog2015}.

The time integration of Eqs. (9)-(12) is performed using the Verlet algorithm provided
by DualSPHysics, where the density, velocity, position, and temperature of particle
$a$ are advanced from time $t^{n}$ to $t^{n+1}=t^{n}+\Delta t$ using the following
steps
\begin{eqnarray}
\rho _{a}^{n+1}&=&\rho _{a}^{n-1}+2\Delta t\left(\frac{d\rho _{a}}{dt}\right)^{n},
\nonumber\\
{\bf v}_{a}^{n+1}&=&{\bf v}_{a}^{n-1}+2\Delta t\left(\frac{d{\bf v}_{a}}{dt}\right)^{n},
\nonumber\\
{\bf x}_{a}^{n+1}&=&{\bf x}_{a}^{n}+\Delta t{\bf v}_{a}^{n}+0.5\Delta t^{2}
\left(\frac{d{\bf x}_{a}}{dt}\right)^{n},\nonumber\\
T_{a}^{n+1}&=&T_{a}^{n-1}+2\Delta t\left(\frac{dT_{a}}{dt}\right)^{n}.
\end{eqnarray}
Numerical coupling of the discrete SPH equations is ensured during the evolution by
alternating the above steps by the form
\begin{eqnarray}
\rho _{a}^{n+1}&=&\rho _{a}^{n}+\Delta t\left(\frac{d\rho _{a}}{dt}\right)^{n},
\nonumber\\
{\bf v}_{a}^{n+1}&=&{\bf v}_{a}^{n}+\Delta t\left(\frac{d{\bf v}_{a}}{dt}\right)^{n},
\nonumber\\
{\bf x}_{a}^{n+1}&=&{\bf x}_{a}^{n}+\Delta t{\bf v}_{a}^{n}+0.5\Delta t^{2}
\left(\frac{d{\bf x}_{a}}{dt}\right)^{n},\nonumber\\
T_{a}^{n+1}&=&T_{a}^{n}+\Delta t\left(\frac{dT_{a}}{dt}\right)^{n}.
\end{eqnarray}

The time step $\Delta t=t^{n+1}-t^{n}$ to ensure stability of the above scheme is
calculated as the minimum between the following timesteps:
\begin{eqnarray}
\Delta t_{f}&=&\min _{a}\left(h|d{\bf v}_{a}/dt|^{-1}\right)^{1/2},\nonumber\\
\Delta t_{v,a}&=&\max _{b}|h{\bf x}_{ab}\cdot {\bf v}_{ab}/({\bf x}_{ab}\cdot {\bf x}_{ab}
+\epsilon ^{2})|\nonumber\\
\Delta t_{cv}&=&\min _{a}\left[h\left(c_{a}+\Delta t_{v,a}\right)^{-1}\right],\\
\Delta t_{e}&=&\min _{a}\left(0.1\rho _{a}c_{p}h^{2}/\kappa\right),\nonumber\\
\Delta t&=&\min\left(\Delta t_{f},\Delta t_{cv},\Delta t_{e}\right),\nonumber
\end{eqnarray}
where $c_{a}$ is the speed of sound of the fluid.

\subsection{Numerical model and boundary conditions}

Equations (1)-(3) are numerically integrated in a two-dimensional domain. A horizontal
cylinder of diameter $D=6$ mm is centered in a square plane of the same size as the square
cross-section of the experimental enclosure of Fig. 3. No-slip boundary conditions are applied
at the walls of the square region (i.e., ${\bf v}_{w}={\bf 0}$) and on the surface of the
circular source pipe (i.e., ${\bf v}_{s}={\bf 0}$) as shown in Fig. 1. Isothermal boundary
conditions are implemented at the walls of the cavity. This is a more accurate representation
of the experimental reality, because it takes into account the finite flux of heat through
the walls. Initially, the square region is filled with 508308 uniformly spaced particles
at rest, where the fluid is represented by particles that carry the properties of glycerin
at $20^{\circ}$C, i.e., $\rho =1264$ kg m$^{-3}$, $c_{p}=2368$ J kg$^{-1}$ K$^{-1}$,
$\kappa =0.286$ W m$^{-1}$ K$^{-1}$, $\eta =1.519$ kg m$^{-1}$ s$^{-1}$, and
$\beta =5.97\times 10^{-4}$ K$^{-1}$. These quantities yield the Prandtl number value
${\rm Pr}=12671.05$. With this spatial resolution the inter-particle separation distance is
0.7 mm in both the $x$- and $y$-directions, while the smoothing length is set to
$\approx 0.99$ mm. Incompressibility of the fluid is guaranteed by setting
$c_{0}=50$ m s$^{-1}$ in Eq. (7).

No-slip boundary conditions are implemented at the walls of the vessel using the method
of dynamic boundary particles developed by Crespo {\it et al.} \cite{Crespo2007}. In
this method, a linear distribution of uniformly-spaced particles is placed at the walls
of the enclosure, with separations of $\approx h/1.42$. A second line of uniformly-distributed
particles is placed outside the computational domain at distances $\Delta x/2$ and $\Delta y/2$
from the wall particles so that they are arranged in a staggered grid. This external particles
are used to cope with the problem of kernel deficiency
outside the computational domain. The wall particles are updated using the same loop as
the inner fluid particles and so they are forced to satisfy Eqs. (7) and (9). However,
they are not allowed to move according to Eqs. (10) and (12) so that their initial
positions and velocities (${\bf v}_{w}={\bf 0}$) remain unchanged in time. In this way,
the presence of the wall is modelled by means of a repulsive force, which is derived
from the source term of the momentum Eq. (10) and includes the effects of compressional,
viscous, and gravitational forces. This force is exerted by the wall particles on the
approaching fluid particles only when the latter get closer than a distance $d=2h$ from
the wall. Across the wall, a particle $a^{\prime}$ is assigned a mass $m_{a^{\prime}}=m_{w}$,
and a density $\rho _{a^{\prime}}=\rho _{w}$, while in order to assure Neumann boundary
conditions for the pressure $p_{a^{\prime}}=p_{w}$ (where the subscript $w$ indicates wall
properties). With a number of 508308 fluid particles, 2917 linearly distributed particles are
needed outside the computational domain, giving a total number of 511225 SPH particles.
Isothermal wall boundary conditions are specified by setting $T_{a^{\prime}}=2T_{w}-T_{a}$
\cite{Szwec2011}, where the wall temperature, $T_{w}$, is initially assumed to be in thermal
equilibrium with the glycerin at $20^{\circ}$C and $T_{a}$ is the temperature of fluid
particle $a$ when it is at a distance $<2h$ from the wall.

The circle at the center of the square cavity in Fig. 1, representing the surface of the 
cylindrical pipe, is also made of fixed boundary particles carrying a a temperature $T_{s}$.
Six model calculations are considered where the value of $T_{s}$ is varied above and below 
the initial fluid temperature $T_{0}=20^{\circ}$C (293.15 K).

\section{Experimental methods}

In order to validate the results provided by the SPH simulations, detailed measurements
of the natural convection flow were obtained experimentally.

\subsection{Test apparatus}

A slender acrylic enclosure of height $H=0.5$ m, length $L=0.5$ m, and width $W=0.025$ m,
filled with glycerin (Pr$=12671.05$ at $20^{\circ}$C) as the working fluid, was employed
for the experiments. The walls of the enclosure thus formed were 6 mm thick to prevent
possible optical distortions (Figs. 2 and 3). A horizontal pipe with a circular cross
section of diameter $D=0.006$ m spanned the width of the rectangular cavity. Its symmetry
axis passed through the points ($H/2$,$L/2$) located on the front and back walls of the
cavity, as shown in Fig. 3. The pipe was part of a closed circuit which allowed the
circulation of water. The temperature of water was tightly controlled by means of a thermal
bath within an accuracy interval of $\pm 0.001^{\circ}$C. Such an arrangement enabled an
approximately uniform and steady temperature difference ($\Delta T$) between the horizontal
pipe and the glycerin within the cavity. The temperature difference was measured every four
seconds with five type $T$ thermocouples (Tc), where four of them were installed on the back
wall with a perpendicular orientation with respect to the pipe, while the other one was attached
directly to the pipe's wall (see Fig. 3). The measurements from the thermocouples were acquired
by means of a Digi Sense Scanning Thermometer Cole Parmer (Mod. 92000-00) temperature scanner.

The hydrodynamical characteristics of the flow were determined with a Particle Image
Velocimetry (PIV) technique, which enabled an accurate spatial and temporal resolution
of the velocity field. All measurements were carried out on a plane, perpendicular to the
axis of symmetry of the pipe, generated with a laser light sheet. A 1200 mJ pulsating Nd:
YAG Laser, with a pulse duration of 4 ms and a wavelength varying from 1064 nm to 582 nm,
was used in this study. Images were recorded with a Hisense MKll camera (with an image
resolution of $1260\times 1024$ pixels) equipped with a 60 mm Nikon lens and a 532 nm
filter. The glycerin was seeded with polyamide (spherical) tracer particles, with a mean
diameter of 20 $\mu$m and relative density $\rho _{\rm glycerin}/\rho _{\rm water}\approx 1$.
The Dantec Dynamics Studio software was employed to process the acquired images.

\subsection{Experimental procedure}

The cell was filled with a prepared mixture of glycerin and tracer particles. The seeding
concentration was optimized to yield a high statistical correlation for the displacement
of particles in the domain. To achieve a detection probability of at least 90\%, the mean
number of tracers per interrogation window was set to be greater than 15 \cite{Keane1990}.
The laser intensity and exposure were adjusted to generate sufficient light scattering by
the tracers and guarantee a good signal quality and flow traceability. The cameras were
focused and image calibration was attained using a ruler placed in the center plane
illuminated by the laser source. In the experiments, the temperature difference was defined as
$\Delta T=T_{\rm bath}-T_{\rm glycerin}$. As it is common practice in most natural convection
experiments, we first considered heating the horizontal pipe with $\Delta T$ values between
0.1 and 10 K. We then considered the opposite case, where the pipe was cooled with temperature
differences in the interval ($-10$ K,$-0.1$ K). Each experiment was repeated three times to
check its reproducibility.

Initially, a two-valve bypass prevented the flow of water inside the pipe. The water was thus
forced to circulate through the bath, until it acquired a state of thermal equilibrium at
$T_{\rm bath}$. At this point, both valves were opened to establish the flow inside the pipe
and initiate the heat transfer process with the glycerin. All experiments were performed at
a room temperature of $20^{\circ}$C.

\subsection{Reproducibility of the experimental measurements}

The experimental reproducibility was assessed by repeating all experiments three times.
Figure 4 shows steady-state profiles of the velocity projected along the line contained in 
the $z=0$ plane that passes through the center of the heat source. In particular, the plot
shows the velocity as a function of height for the case when $\Delta T=10$ K and Ra $=242$
for three different measurements starting from identical conditions, namely
Test 1, Test 2, and Test 3. The vertical velocity component is very sensitive to small changes 
in the initial conditions and, therefore it is a convenient parameter to measure the 
reproducibility of the experiments. The form of the profiles is consistent with a numerical 
prediction by Lauriat and Desrayaud \cite{Lauriat1994} for laminar buoyancy-induced flows 
above a horizontal thin cylinder immersed in an air-filled vessel. A non-zero velocity exists 
close to the cylindrical surface as expected when the pipe behaves as a heat source. In terms of 
the vertical velocity component the differences between the three experimental results are 
small throughout the entire profiles. The maximum velocity for Test 1 is
$v_{\rm max}\approx 5.24\times 10^{-4}$ m s$^{-1}$ and occurs at a height $y\approx 0.344$
m above the source. For comparison, the maximum velocities for Tests 2 and 3 are,
respectively, $v_{\rm max}\approx 5.22\times 10^{-4}$ m s$^{-1}$ (at height
$y\approx 0.362$ m) and $\approx 5.25\times 10^{-4}$ m s$^{-1}$ (at height
$y\approx 0.350$ m).

We may then calculate the differences between the profiles of Test 2 and 3, with respect
to Test 1, by utilizing the root-mean-square error (RMSE) as a metric. The approximate
errors are $\approx 2.7\times 10^{-5}$ m s$^{-1}$ between the profiles 1 and 2 and
$\approx 2.6\times 10^{-5}$ m s$^{-1}$ between the profiles 1 and 3. These results
demonstrate that the experiments are reproducible with good accuracy.

\section{Validation of the SPH simulations}

\subsection{Convergence to the experimental data}

From the numerical point of view, a particle-number independence test was necessary to
determine the quality of the numerical solution as the number of SPH particles is
increased. The experimental profile corresponding to Test 1 was selected as a validation
standard. Figure 5 illustrates the quality of the numerical results as the number of 
particles is increased. The initial inter-particle separations at four
different resolutions are 1.55 mm for $N=104976$ (dotted line), 1 mm for $N=251001$ 
(dot-dashed line), 0.7 mm for $N=511225$ (dashed line), and 0.5 mm for $N=10002001$ (solid
line). The numerical profiles show an asympototic tendency to globally converge 
as the number of particles is increased. In terms of the 
RMSE metric the errors between the numerical and the experimental data for Test 1 are 
approximately $5.63\times 10^{-5}$ m s$^{-1}$ (for $N=104976$), $3.25\times 10^{-5}$ m 
s$^{-1}$ (for $N=251001$), $3.01\times 10^{-5}$ m s$^{-1}$ (for $N=511225$), and
$2.83\times 10^{-5}$ m s$^{-1}$ (for $N=10002001$). On the other hand, the maximum velocities 
and heights are $v_{\rm max}\approx 5.10\times 10^{-4}$ m s$^{-1}$, $y\approx 0.350$ m (for 
$N=104976$), $v_{\rm max}\approx 5.15\times 10^{-4}$ m s$^{-1}$, $y\approx 0.355$ m (for 
$N=251001$), $v_{\rm max}\approx 5.20\times 10^{-4}$ m s$^{-1}$, $y\approx 0.365$ m (for 
$N=251001$), and $v_{\rm max}\approx 5.21\times 10^{-4}$ m s$^{-1}$, $y\approx 0.365$ m 
(for $N=10002001$). The curves for $N=511225$ and $N=10002001$ are seen to overlap with a 
RMSE deviation of $5.05\times 10^{-6}$ m s$^{-1}$. On the basis of these results, a spatial 
resolution resulting from $N=511225$ particles was adopted for all subsequent simulations 
conducted in the present study. At this spatial resolution, the numerical errors between 
the numerical and experimental profiles are observed to fall within the experimental 
uncertainty (see Fig. 4). The robustness of the method is further confirmed by the fact 
that these errors decay faster than the characteristic timescale of the heat diffusion 
process.

\subsection{Lid-driven cavity}

The SPH scheme is further validated against the lid-driven cavity test. This test has 
been used as a benchmark for convection heat transfer problems (see Hopp-Hirschler {\it et 
al.} \cite{Hopp2018a} and references therein). The test consists of a circular, isothermal
flow which sets in within a square cavity as a result of a constant velocity on the top wall 
of the cavity. A schematic of the cavity is shown in Fig. 6. We use the same parameters
as in Hopp-Hirschler {\it et al.} \cite{Hopp2018a}, i.e., the side length of the cavity is 
$L=1$ mm, while the fluid density and dynamic viscosity are $\rho =1000$ kg m$^{-3}$ and
$\eta =0.01$ Pa$\cdot$s, respectively. Three different calculations are considered for 
varying velocity of the top wall of the cavity, i.e., $v_{w}=0.1$ m s$^{-1}$, 1 m s$^{-1}$, 
and 10 m s$^{-1}$, corresponding to Reynolds numbers of ${\rm Re}=\rho Lv_{w}/\eta =10$, 
100, and 1000, respectively. No-slip boundary conditions are applied at the walls of the 
cavity, while Neumann boundary conditions are employed for the pressure at the walls. At 
the bottom corners of the cavity the pressure gradients are either very low or even zero. 
Therefore, in order to avoid the onset of spurious pressures at the bottom corners, the 
reference pressure there is set to zero \cite{Hopp2018a}.

Figure 7 shows the dimensionless $y$-component of the velocity, $v_{y}/v_{w}$, as a 
function of the dimensionless horizontal coordinate position, $x/L$, at $y/L=0.5$ (left
column of plots) and the dimensionless vertical coordinate position, $y/L$, as a function 
of the dimensionless $x$-component of the velocity, $v_{x}/L$, at $x/L=0.5$ (right column
of plots) for ${\rm Re}=10$, 100, and 1000. The SPH profiles at a spatial resolution of 
$N=240\times 240$ particles are compared with the OpenFOAM simulations of Hopp-Hirschler
{\it et al.} \cite{Hopp2018a} using a mesh of $240\times 240$ regular cells. The results
from these simulations are taken as reference solutions for comparison with the SPH
calculations. The velocity profiles are very well reproduced by the present SPH scheme. 
Using as an indicator of convergence the position of the center of the vortex, the relative 
errors of the velocity profiles at this position between the $N=240\times 240$ SPH 
calculations and the reference OpenFOAM solutions are ($\epsilon _{v_{x}}=0.57$\%,
$\epsilon _{v_{y}}=0.49$\%) for ${\rm Re}=10$, ($\epsilon _{v_{x}}=0.25$\%,
$\epsilon _{v_{y}}=0.30$\%) for ${\rm Re}=100$, and ($\epsilon _{v_{x}}=0.59$\%,
$\epsilon _{v_{y}}=0.34$\%) for ${\rm Re}=1000$. For comparison, Hopp-Hirschler {\it et al.} 
reported for this test relative errors of ($\epsilon _{v_{x}}=0.60$\%,
$\epsilon _{v_{y}}=0.42$\%) for ${\rm Re}=10$, ($\epsilon _{v_{x}}=0.24$\%,
$\epsilon _{v_{y}}=0.31$\%) for ${\rm Re}=100$, and ($\epsilon _{v_{x}}=0.60$\%,
$\epsilon _{v_{y}}=0.32$\%) for ${\rm Re}=1000$ between their $N=240\times 240$ SPH results 
and the OpenFOAM solution.

\section{Results}

In a rather global sense, the flow within the enclosure is set in motion under the
action of the buoyant forces, which are induced by the temperature gradients established
between the heat source and the surrounding fluid. When the temperature of the source is
higher than that of the glycerin (i.e., $\Delta T>0$), a vertical ascending flow takes place
in the upper part of the cavity. In contrast, when the temperature difference is
$\Delta T<0$, a descending flow occurs occupying the bottom part of the cavity. Here
we consider values of $\Delta T$ as high as 10 K. However, for values this large the
Boussinesq approximation is still valid because $\beta\Delta T=5.97\times 10^{-3}<1$,
where $\beta =5.97\times 10^{-4}$ K$^{-1}$ for glycerin at $20^{\circ}$C.

\subsection{Flow field visualization}

Flow visualization was carried out by recording the displacement of the tracers
within the cell. The first and third rows of Fig. 8 display the experimental
velocity maps (first row) and their corresponding streamlines (third row) for
$\Delta T=10$ K (${\rm Ra}=242$), $\Delta T=5$ K (${\rm Ra}=60.8$),
and $\Delta T=1$ K (${\rm Ra}=15.2$). For comparison, the second and fourth rows show
the corresponding fields produced by the SPH simulations for the same temperature differences.
For all three temperature differences, an ascending stationary flow forms, which consists of
a pair of counter-rotating vortices filling the upper part of the cavity just above the
heated cylinder. The size and shape of the vortices depend on the temperature difference
and Rayleigh number. From the last two rows of Fig. 8 it is evident that the vortices
grow in size with the temperature difference. The red regions in the velocity maps
correspond to zones of maximum velocity. In these regions the flow is vertically ascending
(i.e., it moves away from the heat source) and the flow streamlines (experiments) and
velocity vectors (SPH calculations) are closer together. As was previously found by
Cesini {\it et al.} \cite{Cesini1999} using air at atmospheric pressure as the working
fluid and Atmane {\it et al.} \cite{Atmane2003} using water, the flow pattern on the
upper part of the cavity is characteristic of naturally convected heat transfer from a
heated circular cylinder. The circulations expand in the upper part of the cavity,
and as the temperature difference increases, both the buoyancy driven flow pattern
and convection heat transfer increase. While the convective flow is confined in the upper
cavity, heat transfer by conduction prevails at the bottom. Apart from minor details, the
SPH calculations reproduce other relevant features of the experimentally obtained flow
patterns. For instance, with an increase of the Rayleigh number, the density variation
becomes greater, causing a more elongated thermal plume (the yellow and red regions in both
the experimental and SPH velocity maps).
\begin{table}[t]
\caption{RMSE errors between the experimental and the SPH profiles of the
vertical velocity component in Figs. 9 and 12}
\begin{center}
\begin{tabular}{c c c}
\\
\hline
 & $\Delta T=10$ K & $\Delta T=5$ K \\
\hline
$y/H$ & RMSE & RMSE \\
 & (m s$^{-1}$) & (m s$^{-1}$) \\
\hline
0.52 & $3.51\times 10^{-5}$ & $2.86\times 10^{-5}$ \\
0.68 & $4.58\times 10^{-5}$ & $1.99\times 10^{-5}$ \\
0.76 & $3.41\times 10^{-5}$ & $1.95\times 10^{-5}$ \\
0.84 & $2.44\times 10^{-5}$ & $1.12\times 10^{-5}$ \\
0.92 & $2.35\times 10^{-6}$ & $8.52\times 10^{-6}$ \\
\hline
 & $\Delta T=-5$ K & \\
\hline
$y/H$ & RMSE & \\
0.12 & $3.76\times 10^{-5}$ & \\
0.24 & $5.62\times 10^{-5}$ & \\
0.36 & $6.28\times 10^{-5}$ & \\
0.48 & $4.93\times 10^{-5}$ & \\
0.60 & $1.81\times 10^{-5}$ & \\
\hline
\end{tabular}
\end{center}
\end{table}

Figure 9 depicts the profiles of the vertical velocity component, $v$,  at different
heights in the upper part of the cavity for $\Delta T=10$ K (left) and $\Delta T=5$ K
(right); the symbols correspond to the experimental data and the solid lines to the
SPH results. At all heights, the vertical velocity component peaks just above the
cylindrical heat source. The maximum velocities occur close to the source. At a height
$y/H=0.68$ the velocities are $v\approx 5.62\times 10^{-4}$ m s$^{-1}$
(experiment) and $v\approx 5.54\times 10^{-4}$ m s$^{-1}$ (SPH) for $\Delta T=10$ K,
while at a height $y/H=0.52$ the velocities are $v\approx 4.89\times 10^{-4}$ m s$^{-1}$
(experiment) and $v\approx 4.81\times 10^{-4}$ m s$^{-1}$ (SPH) for $\Delta T=5$ K.
At greater heights, the maximum vertical velocity component decreases. Moreover, at
either side of the cylinder the vertical velocity component decreases steeply across the
vortices, reaching values as low as $-0.5\times 10^{-4}$ m s$^{-1}$,
regardless of the height. Because of the circulations, the negative values of the
vertical velocity components indicate an ascending flow near the lateral walls of the
cavity. Again, the agreement between the experimental and SPH velocity profiles of Fig. 9
is measured in terms of the RMSE metric. These values are listed in Table 1. The lowest
errors in both cases always occur when $y/H=0.52$, i.e., close to the heat source. At
greater heights, the RMSE errors increase marginally. However, in all cases the errors are
always less than $5\times 10^{-5}$ m s$^{-1}$.

Figure 10 shows the profiles of the vertical component of the velocity along a vertical
line in the $z=0$ plane passing through the center of the source for temperature
differences of 1, 5, and 10 K. These profiles correspond to the velocity field inside
the red plumes shown in Fig. 8. In all cases, the magnitude of the velocity
achieves maximum values in the inner region of the plume close to the source
($0.6<y/H<0.7$). As the temperature difference increases, the region
of maximum velocity becomes progressively broader. This observation is consistent
with the red plumes of Fig. 8 that become more elongated for higher temperature
differences. A non-zero velocity exists at the heat source ($y/H=0.5$). This feature,
as well as the form of the profiles shown in Fig. 10, are in good agreement with the
predictions of Lauriat and Desrayaud \cite{Lauriat1994}, who reported 2D calculations
for the time-dependent laminar buoyancy-induced flow above a horizontal line heat source
immersed in an air-filled vessel. The SPH profiles match the experimental data with
RMSEs of $\approx 1.4\times 10^{-4}$ m s$^{-1}$ for $\Delta T=10$ and 5 K, and of
$\approx 8.1\times 10^{-5}$ m s$^{-1}$ for $\Delta T=1$ K. Further calculations with
lower spatial resolution (not reported here) show that the agreement deteriorates
particularly around the maximum. Therefore, we expect that by increasing the resolution
above $N=511225$ particles the quality of the results will improve. From Figs. 8 and 9 it
follows that the steady-state flow patterns are not exactly symmetric with respect to
the central vertical axis ($x=0$). Asymptotic flow predictions by Angeli {\it et al.}
\cite{Angeli2008} indicate that steady non-symmetric temperature and velocity fields may
sometimes appear in double-cell circulation patterns for different ratios of the cylinder's
diameter to the cavity's side and low ${\rm Ra}$-values.

When $\Delta T<0$, the cylindrical source is colder than the surrounding glycerin,
thereby producing a vertically descending flow. A pair of counter-rotating vortices
forms again, which expands in the bottom part of the cavity. Figure 11 shows the details
of the flow for $\Delta T=-5$ K (${\rm Ra}=93.8$) and $\Delta T=-1$ K (${\rm Ra}=15.7$).
The counter-rotating vortices are nearly symmetrical and advect glycerin from the bulk
of the fluid located beneath the cylinder. A vertically descending flow with a maximum
velocity of $-4.60\times 10^{-4}$ m s$^{-1}$ (experiment) and $-4.61\times 10^{-4}$ m
s$^{-1}$ (SPH) takes place in the red region within the plume. The fine details of the
flow are reproduced by the SPH simulations, because the velocity vectors closely follow
the experimentally obtained streamlines. When the cooler glycerin reaches the bottom
surface of the cavity, it splits up into two nearly symmetric cells occupying the entire
lower half of the enclosure. This can be clearly observed in the experimental and SPH
velocity maps of Fig. 11.

The profiles of the vertical velocity component at different heights from the bottom
surface of the cavity are depicted in Fig. 12 for the descending flows shown in Fig. 11.
The SPH solution (solid lines) is compared with the experimental data (symbols).
Notice how the descending flow velocity always reaches a maximum just below the source.
At either side of the source, the velocity decays sharply and reverts to positive
values close to the lateral surfaces of the cavity. In contrast to the cases where
$\Delta T>0$, the strength of the flow is more sensitive to the temperature difference.
For instance, when $\Delta T=-1$ K, the inverted peaks become much less pronounced and
the ascending flow near the lateral walls of the cavity is practically non-existent.
Table 1 lists the RMSEs of the SPH solutions along with the experimental data. In all
cases, the errors are always less than $6.28\times 10^{-5}$ m s$^{-1}$.

Another comparison between the SPH and the measured maximum values of the vertical
velocity components is illustrated in Fig. 13, for all positive and negative temperature
differences. Positive values of the velocity correspond to the ascending jet produced by 
a positive temperature difference, while negative values of the velocity are associated 
with negative temperature differences. The numerically predicted maximum velocities are
in reasonably good agreement with the experimentally obtained values. The actual difference
between both curves in terms of the RMSE is $8.41\times 10^{-6}$ m s$^{-1}$.

\section{Conclusions}

Natural convection heat transfer from a horizontal, small-diameter cylinder in a
glycerin-filled slender cavity of square-cross section was investigated experimentally
and numerically with the aid of a weakly compressible smoothed particle hydrodynamics
(WCSPH) scheme. Both
the experiments and the numerical calculations were carried out for positive
($0.1\leq\Delta T\leq 10$ K) and negative ($-10\leq\Delta T\leq -0.1$ K) temperature
differences between the source and the surrounding fluid, corresponding to low Rayleigh
numbers between $\approx 1.18$ and $\approx 242$.

Use of the PIV technique allowed to determine the velocity fields for the convection flow
around the cylindrical source along the vertical centerline normal to the cylinder axis.
For $\Delta T>0$, a pair of counter-rotating vortices is formed, which occupies the upper
part of the cavity above the source and grows in intensity for increasing temperature
differences (or equivalently, increasing values of the Rayleigh number). The opposite is
observed when $\Delta T<0$, where now the two circulation zones are confined in the bottom
part of the cavity, which also grow in intensity as the magnitude of the temperature
difference increases. When $\Delta T>0$, the flow pattern just above the source is
characterized by a plume of vertically ascending flow, while the converse occurs when
$\Delta T<0$, where the flow is now vertically descending below the source. In both
cases, the maximum vertical velocities always occur near the source. The SPH
simulations using the same experimental conditions predicted maximum velocities that are
in very good agreement with the experimentally observed values. The main features of the
flow patterns are qualitatively well reproduced by the numerical calculations. Natural
convection under the present geometrical conditions will be more deeply investigated for
larger magnitudes of the temperature difference and varied widths of the cavity.

\begin{acknowledgment} We are grateful to the reviewers who have raised a number of
suggestions and comments that have improved the content of the manuscript. 
F.A. acknowledges the Fondo Sectorial CONACYT - Secretar\'{\i}a
de Energ\'{\i}a for financial support from the Hidrocarburos Estancias Posdoctorales
Program of Mexico. C.E.A.-R. thanks financial support from CONACYT under the Project No. 
368. C.E.A.-R. is a research fellow commissioned to the University of Guanajuato (under 
Project No. 368). We acknowledge financial support from the European Union's Horizon 2020 
Programme under ENERXICO Project, grant agreement No. 828947, and under the Mexican 
CONACYT-SENER-Hidrocarburos, grant agreement No. B-S-69926. The calculations of this
paper were performed using the supercomputing facilities of Abacus-Laboratorio de
Matem\'atica Aplicada y C\'omputo de Alto Rendimiento of Cinvestav-IPN and the Barcelona
Supercomputing Center (MareNostrum).
\end{acknowledgment}

\bibliographystyle{ms}

\bibliography{ms}

\clearpage

\begin{figure*}
\caption{Schematic of the computational domain and its boundary conditions. The circle
at the center represents the cylindrical heat source.}
\label{fig1}
\end{figure*}

\begin{figure*}
\caption{Elements of the test apparatus.}
\label{fig2}
\end{figure*}

\begin{figure*}
\caption{Schematic drawing of the test cell. The legend P1 indicates the location
of the thermocouple on the pipe surface, which serves to measure the
temperature of the pipe $T_{s}=T_{\rm bath}$, while P2 provides a measure of the
glycerin temperature close to the pipe. The legends P3-P6 indicate the location
of the thermocouples on the right side of the back wall of the vessel, which
provide a measurement of the glycerin temperature away from the heat source.}
\label{fig3}
\end{figure*}

\begin{figure*}
\caption{Experimentally obtained profiles for the vertical component of the velocity
along a vertical line on the $z=0$ plane passing through the center of the heat source
for three experiments starting with identical conditions and $\Delta T=10$ K.}
\label{fig4}
\end{figure*}

\begin{figure*}
\caption{Particle number independence study and comparison with the experimental data
of Test 1 (empty circles) for $N=104976$ (dotted line), $N=251001$ (dot-dashed line),
$N=511225$ (dashed line), and $N=10002001$ (solid line) SPH particles.}
\label{fig5}
\end{figure*}

\begin{figure*}
\caption{Schematical configuration of the lid-driven cavity test problem.}
\label{fig6}
\end{figure*}

\begin{figure*}
\caption{Velocity profiles of the vertical velocity, $v_{y}/v_{w}$, as a function of
$x/L$ at $y/L=0.5$ (left column) and of the horizontal velocity, $v_{x}/v_{w}$, as
a function of $y/L$ at $x/L=0.5$ (right column) as obtained for the lid-driven
cavity test for ${\rm Re}=10$, $100$, and $1000$. The SPH profiles (empty circles)
are compared with the reference solution (solid lines) obtained by Hopp-Hirschler
{\it et al.} \cite{Hopp2018a} using an OpenFOAM solver.} 
\label{fig7}
\end{figure*}

\begin{figure*}
\caption{Experimental velocity maps and streamlines (first and third row) as
compared with the SPH simulations (second and last row) for varying temperature
differences: $\Delta T=10$ K (${\rm Ra}=242$; left column), $\Delta T=5$ K
(${\rm Ra}=60.8$; middle column), and $\Delta T=1$ K (${\rm Ra}=15.2$; right
column).}
\label{fig8}
\end{figure*}

\begin{figure*}
\caption{Profiles of the vertical velocity component $v$ across the full length of
the cavity at different heights from the cylindrical source for: ${\rm Ra}=242$
($\Delta T=10$ K) (left) and ${\rm Ra}=60.8$ ($\Delta T=5$ K) (right). For reference
the heat source is at $x/H=0.5$. The symbols depict the experimentally obtained
profiles, while the solid lines correspond to the SPH profiles.}
\label{fig9}
\end{figure*}

\begin{figure*}
\caption{Profiles of the vertical velocity component $v$ along a vertical line
in the $z=0$ plane passing through the center of the source for temperature
differences of 1, 5, and 10 K between the heat source and the surrounding fluid.
The experimentally obtained profiles (symbols) are compared with those obtained
from the SPH simulations (solid lines).}
\label{fig10}
\end{figure*}

\begin{figure*}
\caption{Experimental velocity map and streamlines (first and third row) as compared
with the SPH simulations (second and last row) for $\Delta T=-5$ K (${\rm Ra}=93.8$;
left column) and $\Delta T=-1$ K (${\rm Ra}=15.7$; right column).}
\label{fig11}
\end{figure*}

\begin{figure*}
\caption{Profiles of the vertical velocity component $v$ across the full length
of the cell at different heights from the bottom surface of the cell below the
cylindrical source for $\Delta T=-5$ K (${\rm Ra}=93.8$). The experimentally
obtained profiles (symbols) are compared with those obtained from the SPH simulations
(solid lines). For reference the cylindrical source is at $x/L=0.5$.}
\label{fig12}
\end{figure*}

\begin{figure*}
\caption{Comparison between the experimentally and numerically obtained maximum
values of the vertical velocity component for all temperature differences
considered.}
\label{fig13}
\end{figure*}

\clearpage

\begin{figure*}
\centerline{\epsfig{figure=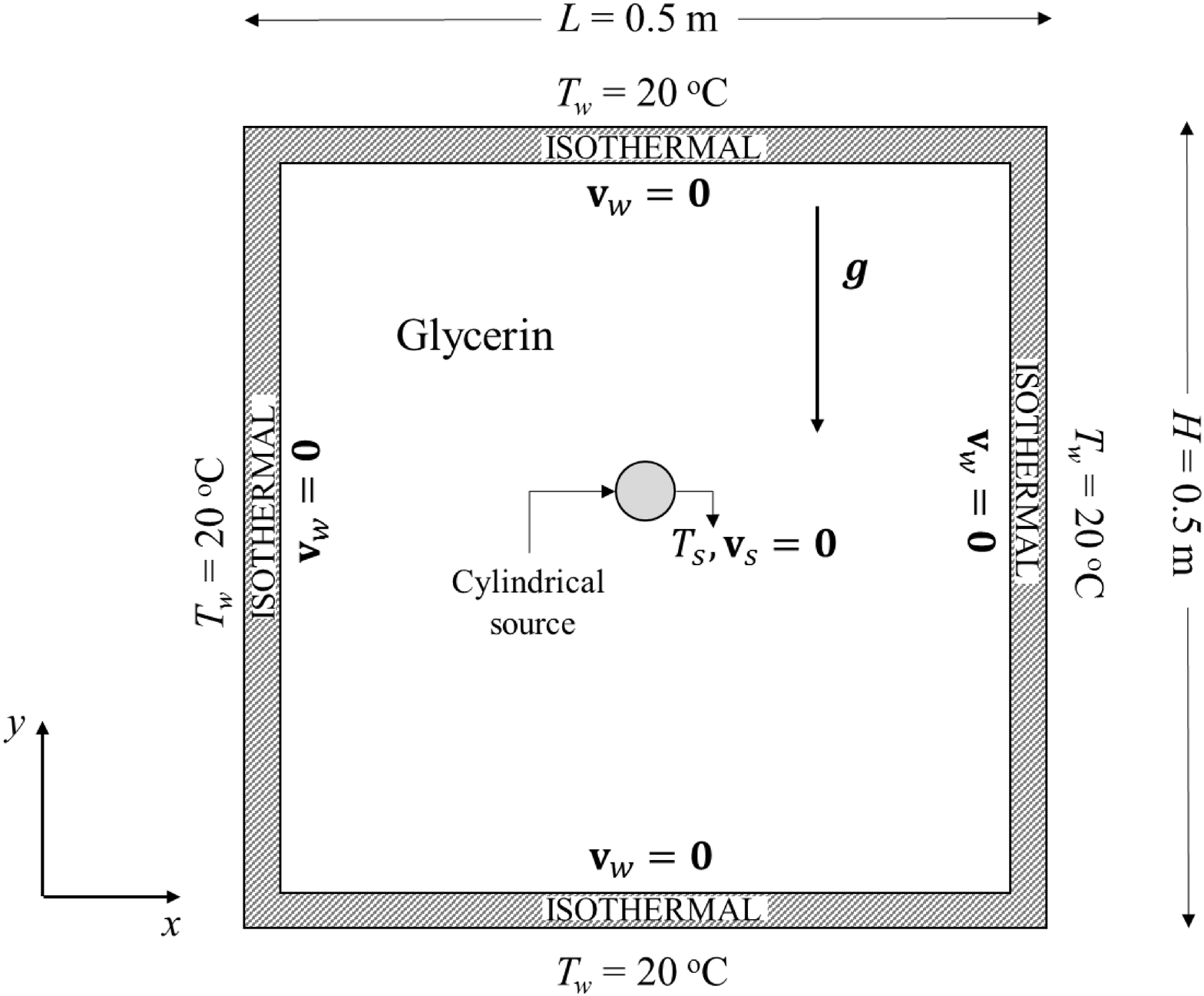,width=5.5in}}
\label{fig1}
FIGURE 1
\end{figure*}

\clearpage

\begin{figure*}
\centerline{\epsfig{figure=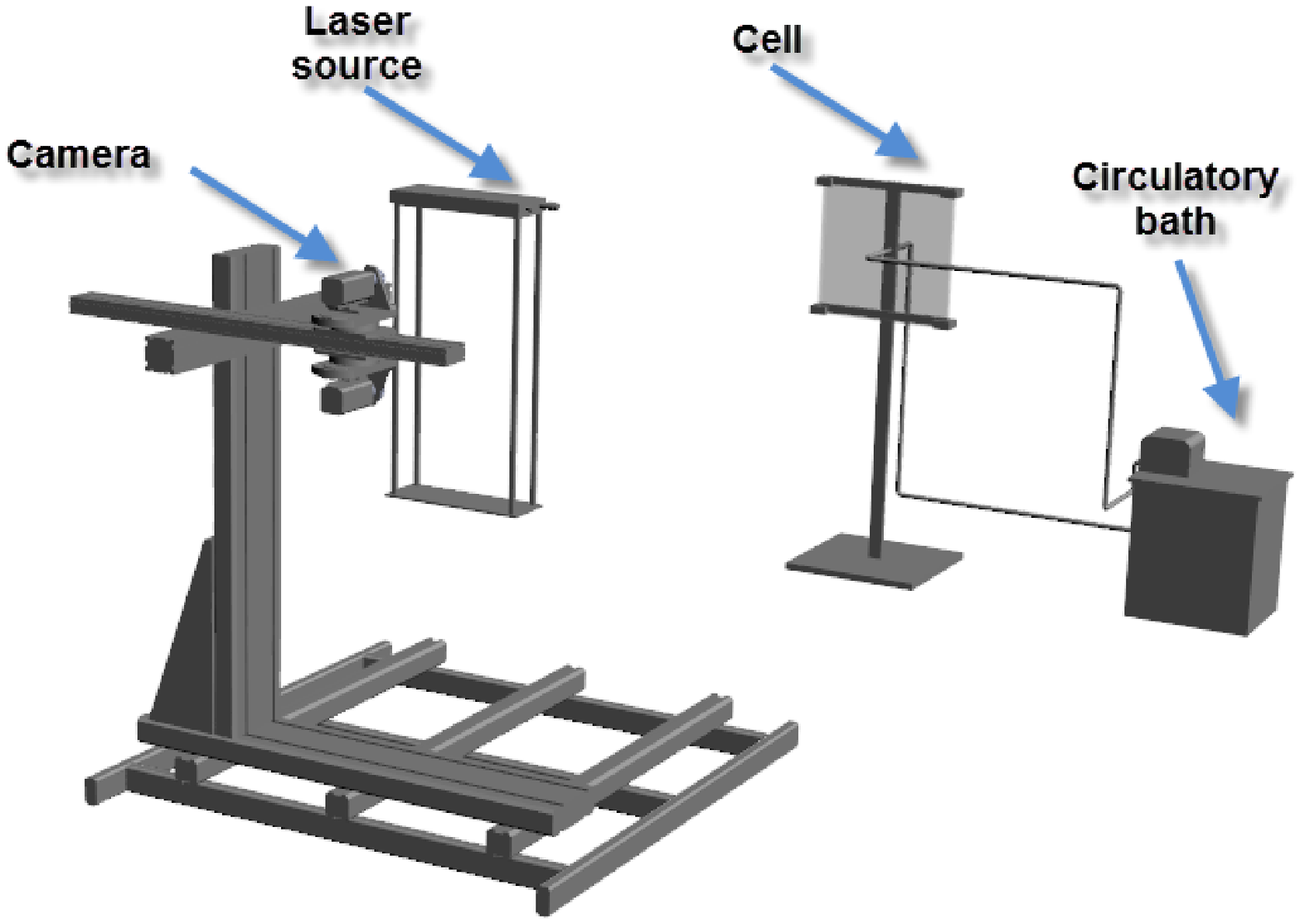,width=5.45in}}
\label{fig2}
FIGURE 2
\end{figure*}

\clearpage

\begin{figure*}
\centerline{\epsfig{figure=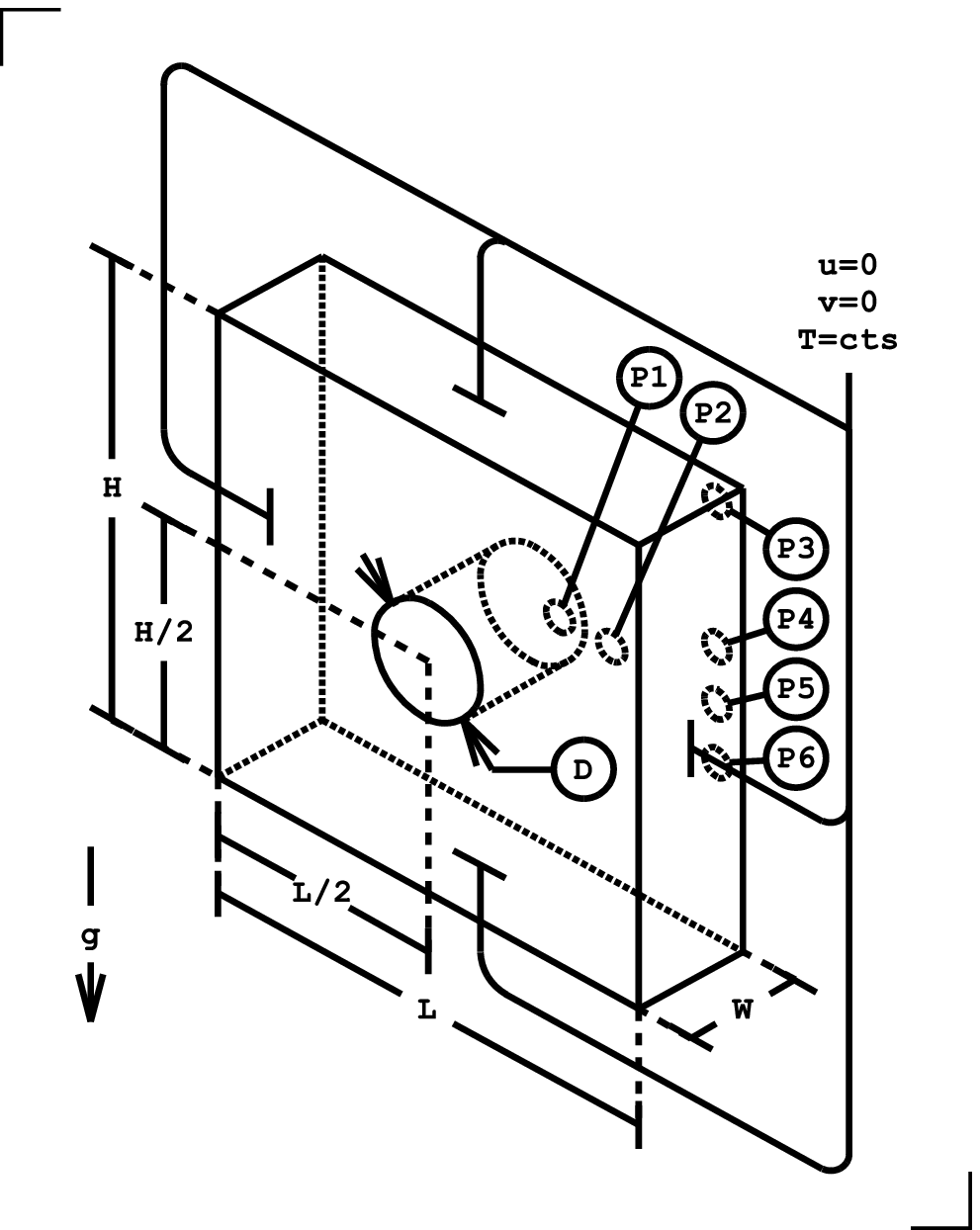,width=3.85in}}
\label{fig3}
FIGURE 3
\end{figure*}

\clearpage

\begin{figure*}
\centerline{\epsfig{figure=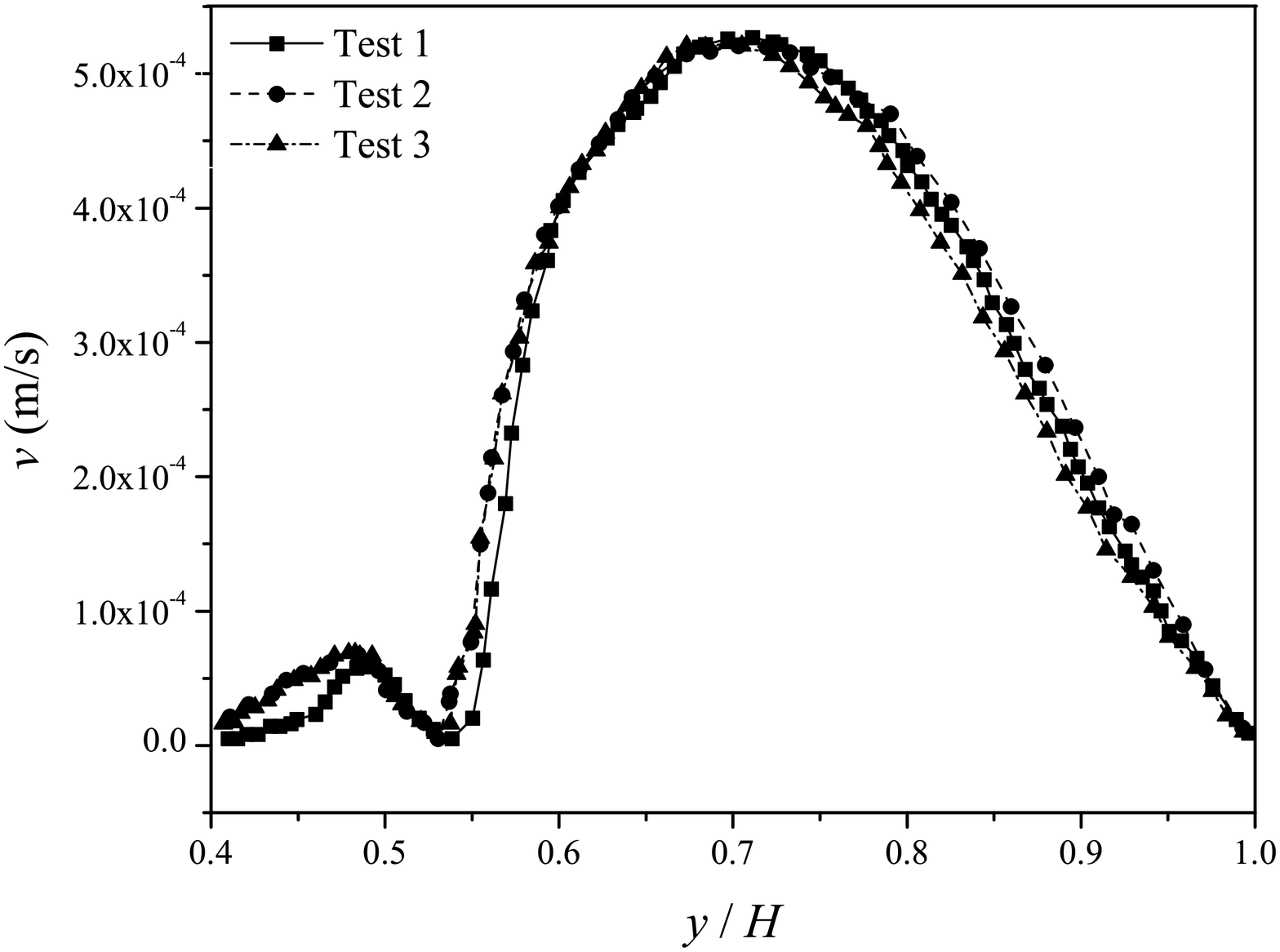,width=5.45in}}
\label{fig4}
FIGURE 4
\end{figure*}

\clearpage

\begin{figure*}
\centerline{\epsfig{figure=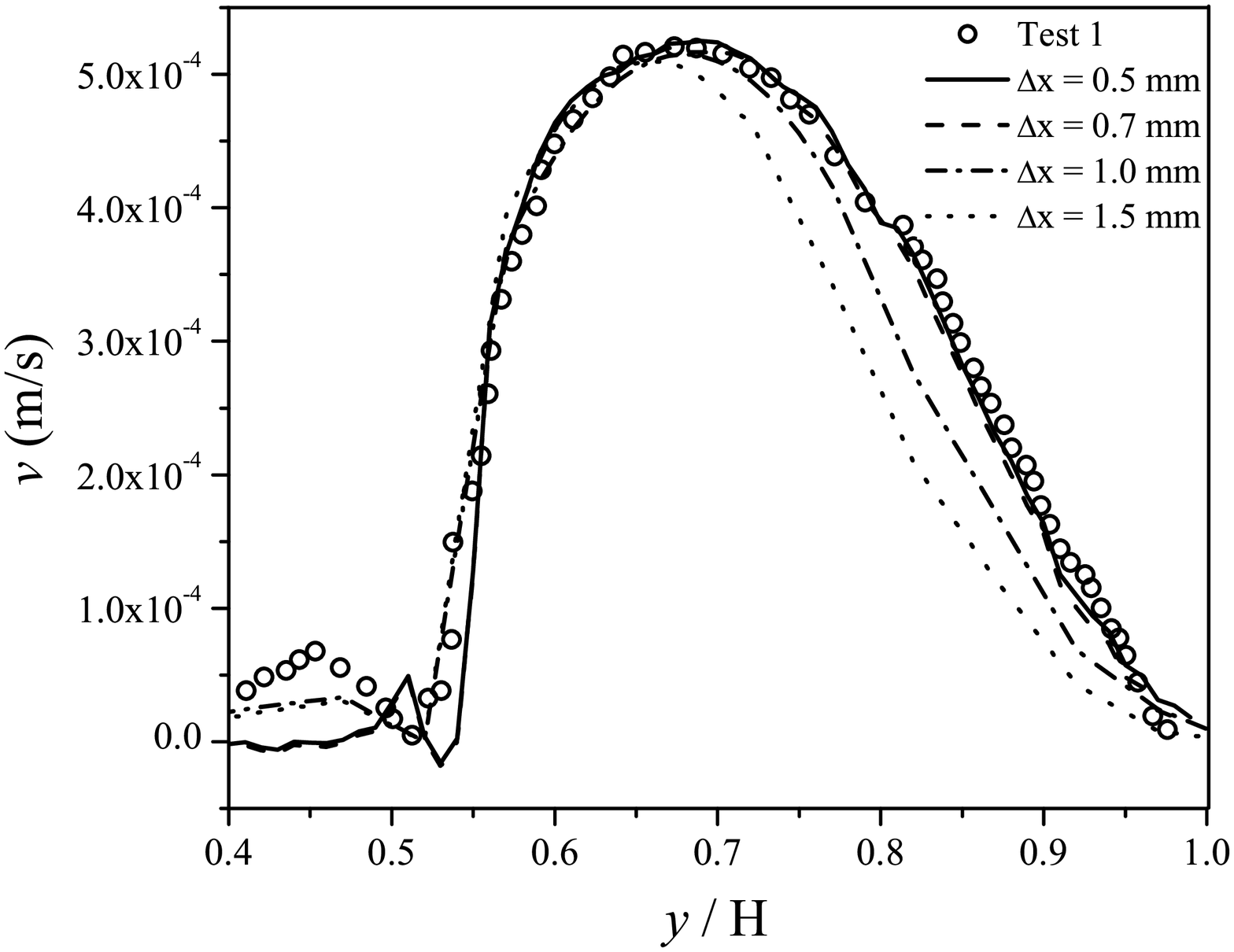,width=5.45in}}
\label{fig5}
FIGURE 5
\end{figure*}

\clearpage

\begin{figure*}
\centerline{\epsfig{figure=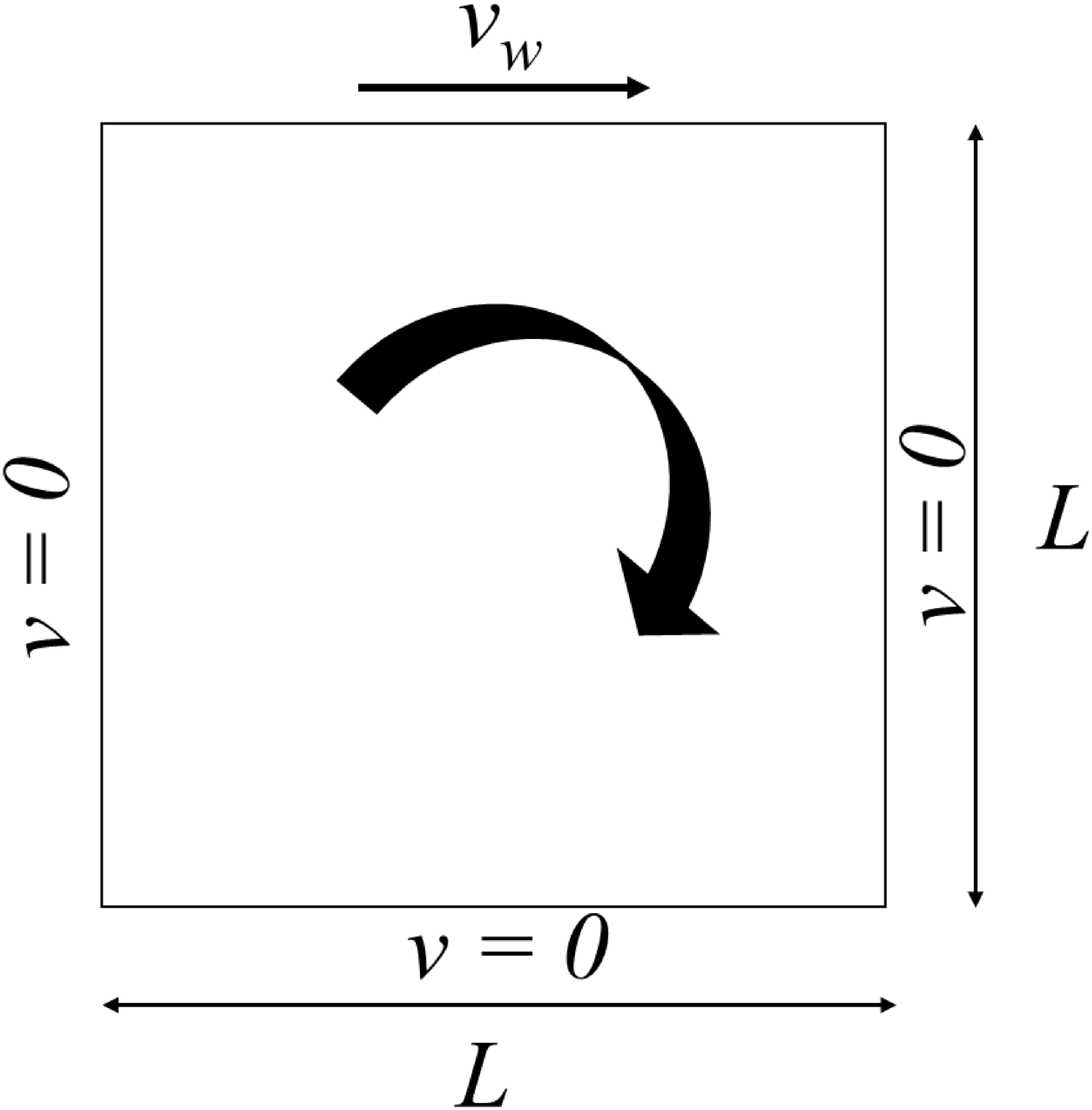,width=5.55in}}
\label{fig6}
FIGURE 6
\end{figure*}

\clearpage

\begin{figure*}
\centerline{\epsfig{figure=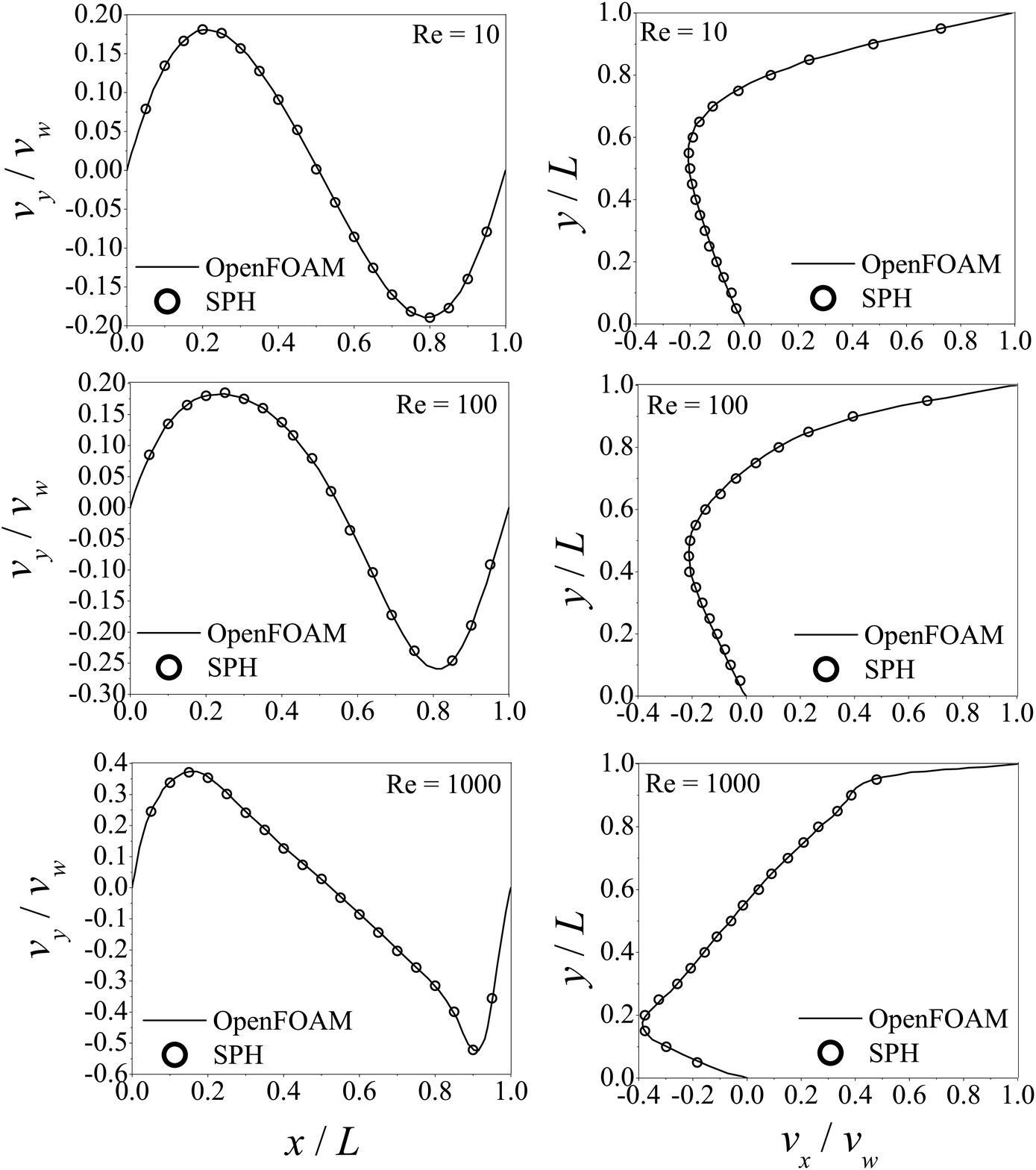,width=5.0in}}
\label{fig7}
FIGURE 7
\end{figure*}

\clearpage

\begin{figure*}
\centerline{\epsfig{figure=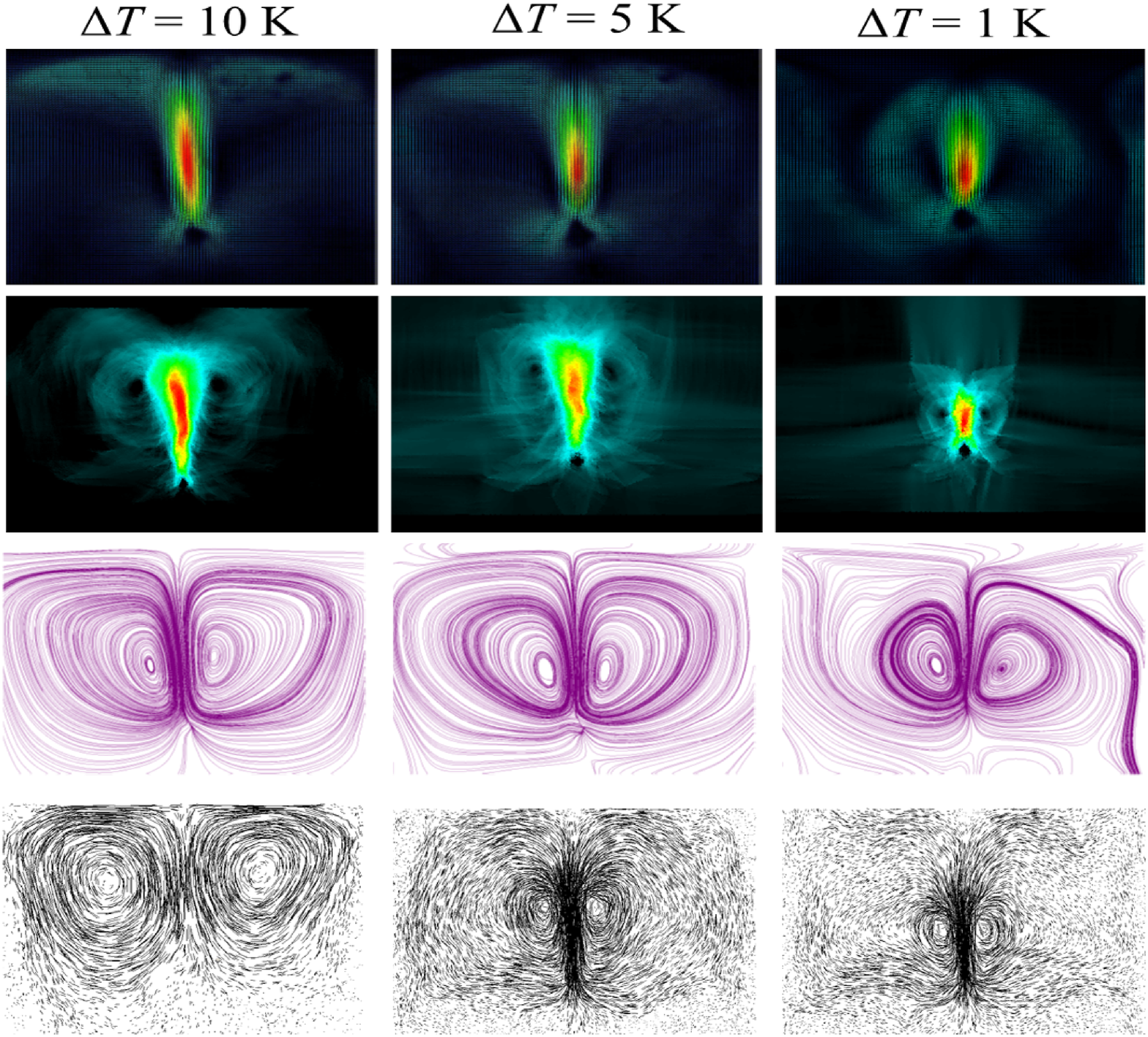,width=6.0in}}
\label{fig8}
FIGURE 8
\end{figure*}

\clearpage

\begin{figure*}
\centerline{\epsfig{figure=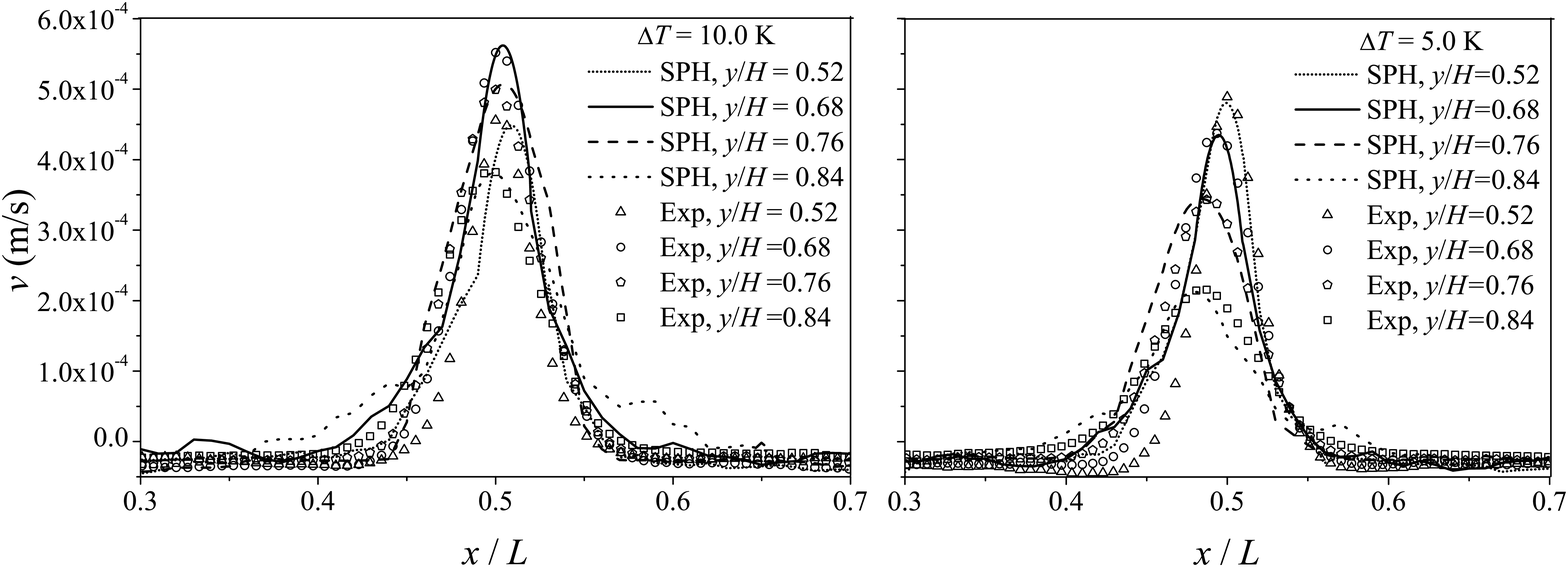,width=5.0in}}
\label{fig9}
FIGURE 9
\end{figure*}

\clearpage

\begin{figure*}
\centerline{\epsfig{figure=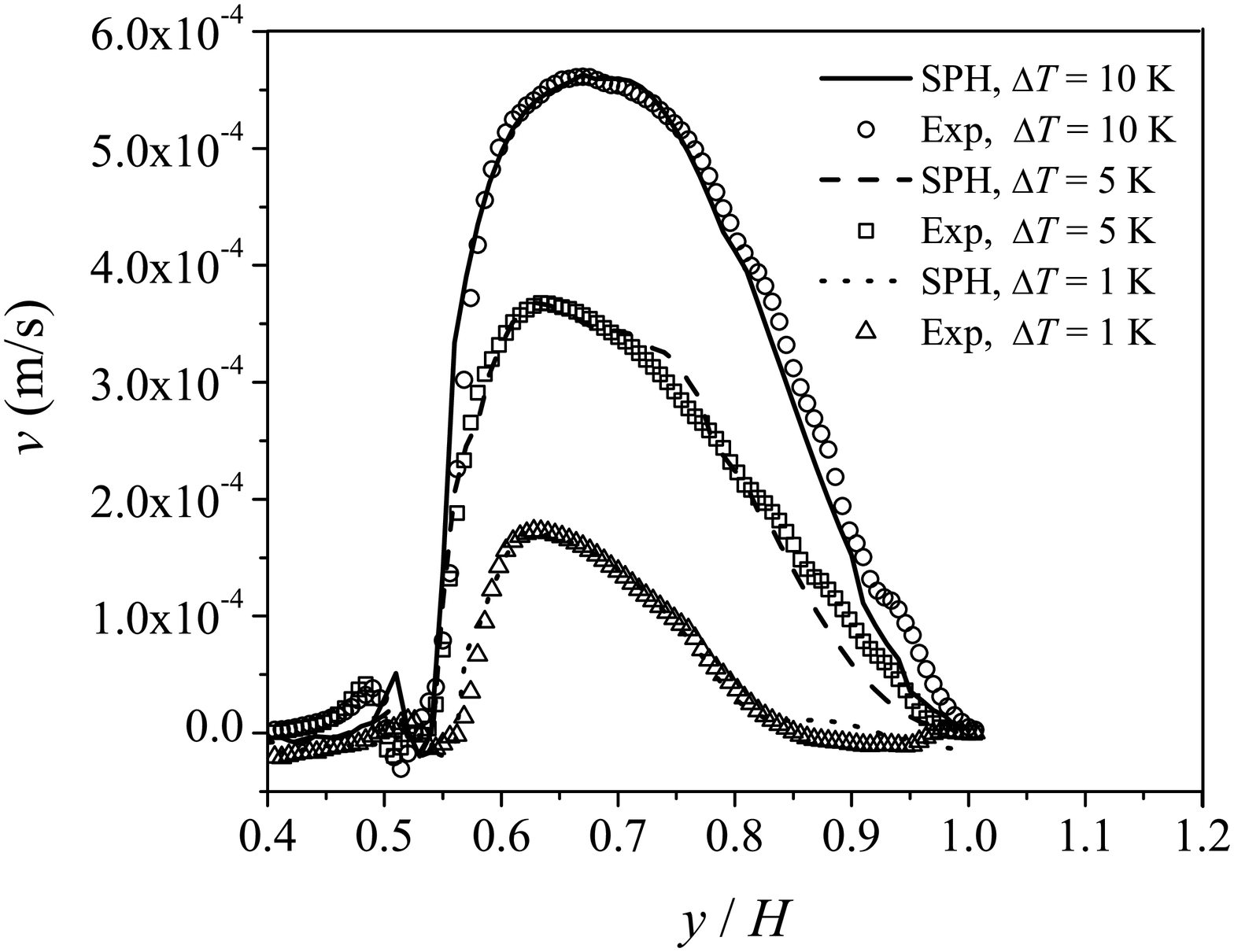,width=5.45in}}
\label{fig10}
FIGURE 10
\end{figure*}

\clearpage

\begin{figure*}
\centerline{\epsfig{figure=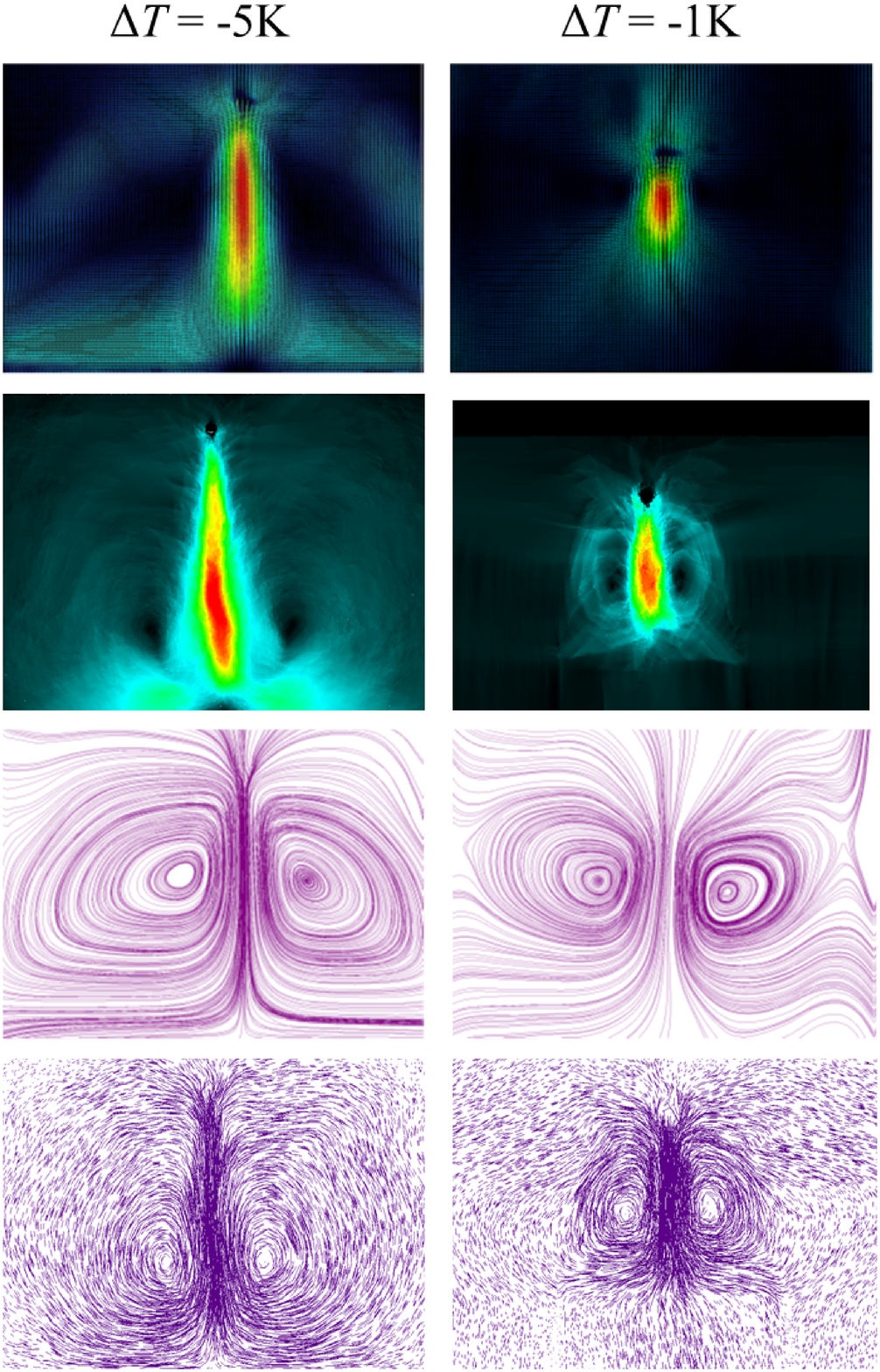,width=6.0in}}
\label{fig11}
FIGURE 11
\end{figure*}

\clearpage

\begin{figure*}
\centerline{\epsfig{figure=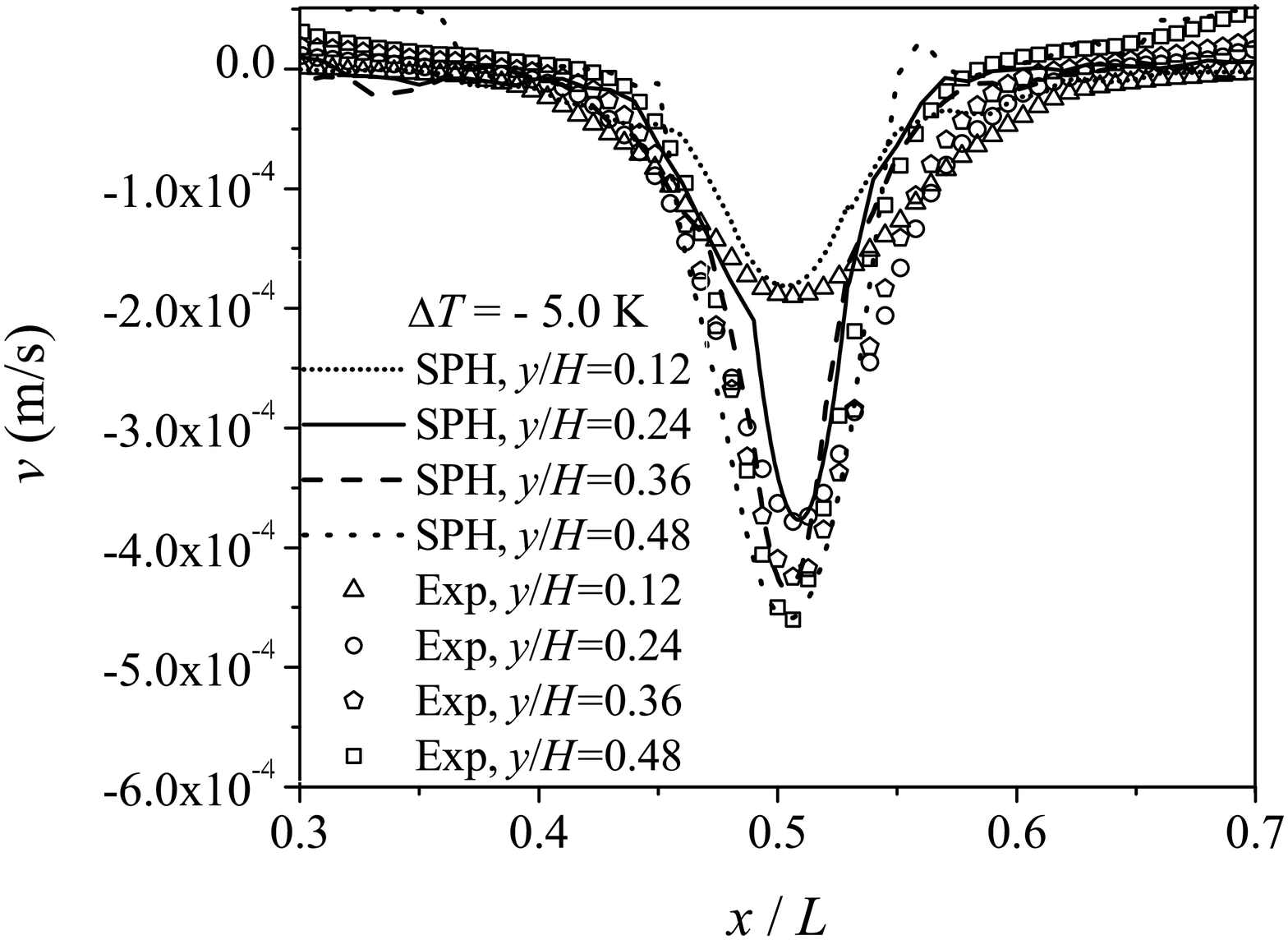,width=4.45in}}
\label{fig12}
FIGURE 12
\end{figure*}

\clearpage

\begin{figure*}
\centerline{\epsfig{figure=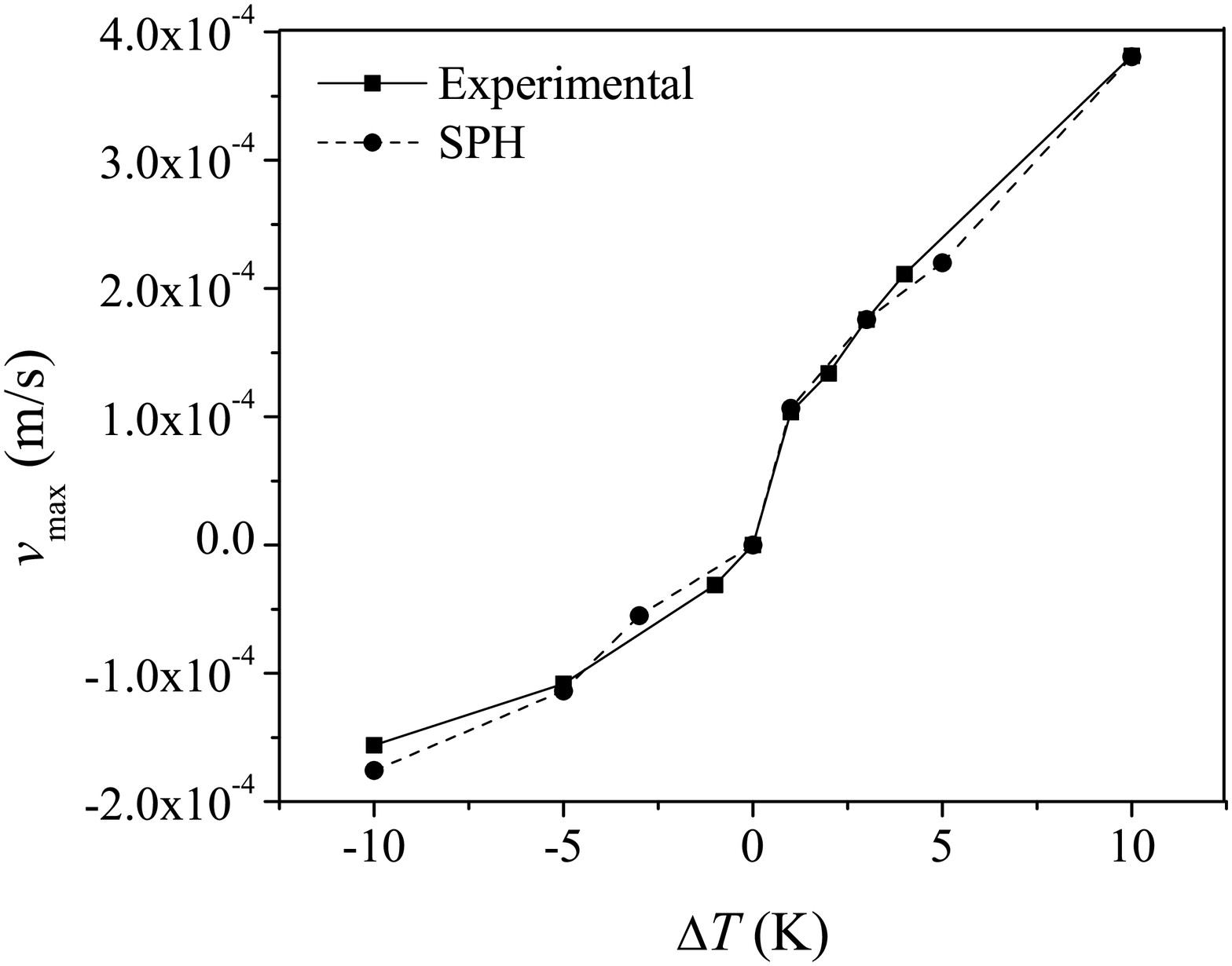,width=5.0in}}
\label{fig13}
FIGURE 13
\end{figure*}

\end{document}